# Beyond the Boundaries of Open, Closed and Pirate Archives: Lessons from a Hybrid Approach

Prodromos Tsiavos, London School of Economics, UK

Petros Stefaneas, National Technical University of Athens, Greece

The creation of open archives i.e. archives where access is regulated by open licensing models (content, source, data), should be seen as part of a broader socio-economic phenomenon that finds legal expression in specific organizational and technical formats.This paper examines the origins and main characteristics of the open archives phenomenon. We investigate the extent to which different models of production of economic or social value can be expressed in different forms of licensing in the context of open archives. Through this process, we assess the extent to which the digital archive is moving towards providing access that is deeper (meaning, that offers more access rights) and wider (in the sense that most of the information given is in open content licensing) or face a gradual stratification and polarization of the content. Such stratification entails the emergence of two types of content: content to which access is extremely limited and content to which access remains completely open. This differentiation between classes of content is the result of multiple factors: from purely legislative, administrative and contractual restrictions (e.g. data protection and confidentiality restrictions) to information economics (e.g. peer production) or social (minimum universal access).

We claim that with respect to the access management model, most of the current archiving processes include elements of openness. Usually, this is the result of economic necessity expressed in licensing instruments or organisational arrangements. The viability and the socio-economic importance of the digital archives also contributes to the use of open archiving practices. In such a context, although pure forms of open digital archives may remain an ideal, the reality of hybrid open digital archives is a necessity.

## 1. Introduction

The emergence of digital networked archives has marked the advent of a new era for the archive and its relationship both with memory institutions and content providers. Whereas, archives have traditionally operated as organisations responsible for the preservation and controlled access to content with a limited impact upon its regular, commercial dissemination and exploitation, the digitisation of the archives, first, and their availability over digital networks, subsequently, has substantially changed their role and impact (Bernstein 2008). In a rather ironical fashion, archives seem to be the victims of their own success. We need to further elaborate on this position.

The digitisation of the material contained in archives and -most importantly- their digital cataloguing and curation has made archives more accessible than ever before. This practically means that the user has the ability to access the content of an archive in a much faster and easy way than in the past, to accurately identify the kind of material she wishes to access and to explore similar content in a more effective and efficient way.

For as long as digital archives remained confined within the walls of their traditional institutional role and physical location, their function remained to a great extent similar to the one they had ever since their inception. However, once the archive became available over digital networks, its nature and boundaries have known a substantial transformation and expansion . Once the archival content becomes available over the public internet, a series of social, economic and legal questions emerge (Dietz 1998). For instance, is such an archive in direct competition with the owners of the Intellectual Property Rights over the relevant content? How can personal data be protected and how can policy makers achieve a balance between preservation of

These questions are further amplified by three increasingly important new types of archives, that is, commercial, open and pirate archives.

Commercial archives have increased in importance since they represent a new mode of exploitation for previously not widely available material. Commercial archives seem to compete with traditional archives and are possibly disincentivised to make their material available to an archive because of negative potential implications for their business model (Creative Archive Licence Group 2006). This becomes very relevant in cases where the archive holds valuable information sources, such as audiovisual material, newspapers or recordings of various kinds. Interestingly, even when the material is not protected under copyright any more may still have access controls based on physical property and the question to which such technical and contractual means may be used in order to restrict and control access is one of the question explored in this paper.

Open archives are the ones where the content is made available through an open licensing scheme, i.e. when the downloading, copying, further use and dissemination of the content in its original or altered form, remains free of any restrictions. In practice, open archives are placed in a spectrum of openness with respect both to the content and its meta-date. Such spectrum covers from full openness, where no restrictions are posed to either content or meta-data with respect to their potential uses. One aspect of openness also includes the ability of the user to add content or meta-data which are then shared with different degrees of openness (Samis 2008).

Pirate archiving is a trend appearing along the emergence of closed peer-to-peer networks. These make extensive use of content for which copyrights have not been cleared, however, the curation of the material is being done by the users of the network and is often superior to the quality of documentation found in regular archives (Bansal, Keller et al. 2006).

This paper explores these three types of archives and the ways in which their features could be combined in order to design a national archival policy for countries that are not net exporters of Intellectual Property Rights, such as Greece, in order to make the most out of the use of open archives. We argue that while open archives in their pure form are not always easy to create or sustain, primarily due to the ways in which the existing legal system operates, it is possible to create hybrid forms of archives with variable degrees of openness that could contribute to the cultivation of an ecology of open archives.

## 2. Research Design and Methodology

In order to explore the ways in which hybrid archives operate, we need to use an analytic tool that explores the way in which value of different types is produced in cases of different archives. The value is not necessarily monetary, it could be e.g. social or other. The value is produced through the flows of content or data which is accordingly regulated by technological or legal means.

2.1 Basic concepts

The methodology employed in this report is based on the identification and analysis of three basic variables that appear in each of the case studies. These variables are as follows:

a) Value

b) Content

c) Rights

Value, content and rights are closely interrelated and it is useful to trace their relationship, as it sets the management framework for any e-content project (Young 2005; Pasquale 2006). However, they need to be kept analytically separate and examined in juxtaposition to each other:

\*   The flow of content produces value: eg when a user downloads a digitised sound recording, the user gains value in terms of knowledge and the public-sector organisation increases the visibility of its collection and hence its cultural value

\*   The flow of content is regulated by the rights existing on it: eg when a work is licensed under a Creative Commons (Lessig 2007) Attribution licence, it may be freely exchanged between users provided they make reference to the author of the work[1]

\*   The flows of content and rights do not follow the same path: eg in the case of User Generated Content (UGC) that resides in a repository and is licensed under a Creative Commons licence, the content flows from the repository to the user, whereas the licence (rights) flows from the user that has authored the content to the one that uses it

This methodology features:

a)   A series of steps to be followed in order to trace flows of value, rights and content in any project. These constitute an analytical framework that may be replicated and employed in any project involving management of rights protected content for the production value

b)   The specific process and rationale of data selection, collection and analysis followed in this project

2.2 Value

Gaining best value from the investment that has been made in the production of publicly funded e-content is among the core objectives of all organisations being studied here. Such value is not necessarily monetary nor of a single type. Different stakeholders have different perceptions of value and the identification of types of value is the first step for achieving any project's objectives (Dyson 1995). Each of the projects presented in this report seeks to achieve a set of objectives that are in turn served by values of variable type that flow into and out of the project. The identification of different types of value presupposes an understanding of the stakeholders and the key objectives of each project.

2.3 Content

There are various types of content that are circulated within the boundaries of a particular project or could potentially flow across different projects. One way of classifying electronic content is on the basis of its source. Three categorisations are made on that basis:[2]

---

[1] See eg Creative Commons Attribution licence, 'Legal Code', Unported, Section 4b, http://creativecommons.org/licenses/by/3.0/legalcode

[2] The categorisation is made from the perspective of the organisation that obtains, produces, hosts and makes content available. The difference between UGC and third-party content is that the former is in most cases produced by individuals that are non-professionals and hence may require different treatment (eg quality testing, filtering etc) compared to content produced by organisations or professionals.

a) User-generated content

b) In-house produced content

c) Third-party content, ie content produced by organisations other than the one hosting it

Each of the aforementioned types of content has different trajectories of flow:

a)   User-generated content tends to flow in a circular form: the content flows from the user to the organisation that manages the project and then again from the organisation to other users. If the material is repurposed then the circle starts again

b)   In-house produced content flows from the organisation that manages the project to: –    Intermediaries that will further disseminate the content to other intermediaries or the end-user –    The end-user

c)   Third-party content flows from the third parties to the organisation managing the project and then back to the user. In the case where only hyperlinks to the third-party content exist, the content flows directly from the third party to the end-user

Another categorisation of the content may be on the basis of its nature. We thus have:

*    Audiovisual works, text (literary works), musical works and sound recordings

*    Raw data and compilations of data

*    Software

*    Multi-layered works: these consist of works comprising multiple layers of other works (eg a multimedia work containing all the aforementioned categories of works, ie audiovisual works, data, text, software)

A final important distinction is between content and metadata, the former referring to the actual works and the latter to information about them. The differentiation is important both because rights may exist on both types and because there are projects that derive their primary value from the production and use of content and others from the production and use of the metadata.

2.4 Permissions and rights

e-Content comprises multiple layers[3] and types[4] of rights that regulate its flow. More specifically, multiple types of rights may exist on a specific work or multiple permissions may be required for its use. For example:

*   Intellectual Property Rights (such as copyrights or trademarks)

*   Permissions to use personal data or information with respect to minors

*   Prior Informed Consent for use of sensitive personal data

To answer this question, we need to explore
(a) the changing role of the archive (cultural and economic)
(b) to present the range of emerging legal issues
(c) to suggest ways in which content/ data and licensing flows are to
be structured in order to mitigate legal risks and maximise the value
produced by an archive

It is important to note that though IPRs are the main focus of this research, the management of certain other types of rights and permissions was also mentioned by some of the case studies. These included the management of confidentiality agreements, obtaining prior informed consent and following data protection legislation, which were considered to be equally if not more important risk-management considerations than the management of IPR.

Multiple layers of rights may exist on what appears to the end-user as a single work. An oral history recording may, for instance, consist of multiple underlying literary works, a performance and the actual sound recordings. Each of these works is awarded by the copyright legislation different sets of moral and economic rights.

These multiple types and layers of rights may well belong to different rights holders, causing significant frictions in the flows of works that are governed by those rights.

In the same way as content flows within and across projects, rights may also be created and transferred between individuals and organisations. Ownership over the physical or digital carrier of a work does not automatically entail ownership of the Intellectual Property Rights or a licence for the distribution or repurposing of e-content. For example, a museum may own a painting but still may not be able to digitise it. Even when the rights owner provides a digitisation licence, this may allow the making of copies only for preservation purposes and not for dissemination to the general public.

---

[3] Eg what appears as a single audio recording may comprise different layers of copyright existing on the literary work, the sound recording and the musical work.

[4] Eg Copyrights, trademarks, personal data.

Rights holders are able to manage their rights by providing different types of licences or permissions allowing licensees to perform specific acts, such as redistributing (sharing) or repurposing content.

2.5 Flows

Identifying different types of value, content and permissions constitutes an important step toward the description of the information blueprint of an organisation, but it lacks the interactive element present in all content-related transactions. It is the flow of value, content and permissions and the relationships between these different streams that provide the complete picture of the operation of the relevant projects (Aigrain 1997).

Focusing on the tracing of flows allows a better understanding of content-related transactions in terms of:

\*      The life cycle of flows and

\*      The association of flows with each other

Overall, the following basic conditions are usually encountered regarding flows:

\*      Flows of value, permissions and content flows are always associated. However, it is not clear whether such associations are beneficial for the objectives of the project or what barriers they face. Flows of permissions and works will inevitably produce some kind of value, but it is important to examine whether such value types are consistent with the project's objectives and the cost of producing such value

\*      Often a project seeks to produce a certain type of value but legal constraints may limit the flows of permissions and hence of works; this may consequently create frictions in the desired flow of value. Such frictions limit or cancel the flow of works. For example, sound recordings may only be used on site, not making use of the available technological options, or digitised recordings may never be made available. As a result, flows of cultural value with respect to specific types of content may be never materialised

2.6 A life-cycle approach

Tracing the life-cycle of flows of value, content and permissions is instrumental for constructing the blueprint of each of the examined projects. It involves the following steps:

a)     Identification of project objectives and types of value

b)     Identification of layers and types of content and rights and assessment of their documentation process

c)      Tracing the cycle of flows of works and permissions within a project: the flows of works and rights do not always coincide or may follow multiple paths. For example, a library may acquire a licence from a researcher for all the rights on a sound recording, but might only license listening to the work to the end-user. A work may enter the museum in a physical form and be made available in a digital form of variable quality to different groups of users

d)      Tracing cycle of flows of works and permissions across projects: organisations of the broader public sector often need to be able to use each other's content. For example, the BBC Century Share project makes the content of other SCA sponsor organisations available to a wider audience than each individual organisation would be able to disseminate it to

e)      Matching flows of works, permissions and value: different types of value are produced as a result of flows of rights and content

2.7 Key factors to be taken into consideration

In each of the stages we further examine:

a)      Association of funding with access and use policies: a significant portion of the e-content produced or made available by SCA sponsor organisations is publicly funded through grants that set specific conditions regarding its dissemination and use. Such conditions provide the framework for access and use policies that need to be followed by the funded project. For example, as a result of JISC funding, project developers will be required to make their project outputs freely available to Higher and Further Education (HE/FE) communities for educational and non-commercial uses. In such cases users often also acquire a licence to share and repurpose the content. Such licences grant far more extensive rights to users compared to rights granted by commercial organisations.

b)      Risk management strategies: collections normally held by the SCA sponsor organisations present rather complex issues because of the multiple types of content and rights involved, and subsequently the potential for numerous transactions. An analysis of the respective organisations with regards to these transactions on the basis of flows of rights and content, allows for the design of more effective risk-management strategies. Effective risk-mitigation strategies facilitate better flows of content and contribute to an increase of flows of value. Most risk-mitigation strategies are based on the following mechanism:

*       Identification of potential risks

*       Impact assessment

\*   Probability of risks

c)  A balance of inputs/outputs of licences/permissions approach: each project was assessed on the basis of the degree to which it ensured the compatibility of permissions that have been secured from third parties and those which the organisation was furthering allowing access and reuse (the rights' input is equal or greater than the rights' output).

2.8 Data collection and research design

The above approach is applied in the following 9 case studies:

* BlueGreece Torrent Tracker

* BBC Creative Archive

* Internet Archive

* Broadcasting Archives

* British Library Archives

* BBC CenturyShare Project

* British Library Archival Sound Recordings (BL ASR I and II)

* National Library for Health eLearning Object Repository (NLH LOR)

* Great Britai Historical Geographic Information System/ Vision of Britain Through Time

In each of the cases we explore the ways in which this analytical scheme may give us some insight as to how licensing schemes may be used in order to produce different types of value.

## 3. Case Studies

In this section we present a series of examples of archives to highlight the different types of organization, types of material and business models.

3.1 Case One: BlueGreece Torrent Tracker

3.1.1 Background

BlueGreece[5] is a private torrent tracker that has been in operation since early 2005. It numbers 37,683 active torrents and 54,782 members. Its most popular file has been downloaded 17,416 times and there are 441 seeders for the most popular item. While the administrators of the site have issued a statement where they disclaim any responsibility over the content that is exchanged over the servers and there is an explicit statement as to how the users should not use the site to download or use material when they do not have the rights to do so, almost the entirety of the content is copyrighted and not licensed to be re-distributed among the users. Since the site is a tracker and the files are directly exchanged between the members of the tracker, the administrators claim they have no responsibility over any copyright infringement taking place. In practice, the site is used for the illegal sharing of copyrighted content and is hence a pirate archive.

The archival nature of the site is supported by the rich curation of the material, the existence of formal requirements as to how it is to be shared and distributed and the penalties in existence when these rules are not followed. It is also important to note that the types of the content found on the tracker also contains the diamond category under which very well curated or very popular material is placed.

3.1.2 Key content features

* Multiple types of works (audio, video, image, text, data-bases)
* Mostly copyrighted material but also self-published/ end-user material
* Mainly copyrights and related rights/ data-base rights
* Extensive documentation and curation of the content

3.1.3 Value gains

* The main objective of BlueGreece is to provide access to user-collected content to a mainly Greek-speaking audience that do not have access to such material otherwise. While most of the content is copyrighted, this is also a platform, where non-professional users are able to distribute their content.
* The main value produced is cultural, both in the sense of preserving and in the sense of curating content that is culturally significant for a group of users. It would be a fallacy to see the BlueGreece as a means by which copyrighted content is made available to users without paying a fee. The

---

[5] The name of the torrent tracker has been purposefully altered so that is not directly identifiable. This is an action taken in order to ensure that the risk of prosecution for the tracker administrators as a result of this research is reduced.

vibrant forum as well as the extensive use of meta-data and documentation accompanying the torrent URIs is indicative of the main type of value produced by the tracker, that is cultural value. Even when the content is not in the Greek language, as the comments of the users indicate, there is great cultural significance attributed to its use (e.g. it was played by the National Broadcaster in the 1980s or 1990s or it is part of a "retrospective" that the uploader creates). One of the main features of the tracker is the ability to access torrents by reference to the uploader or the group of uploaders.

* Another type of value is the one created through the emergence of groups of "cappers", that is groups of persons dedicated to the digital recording of TV shows or sports events or radio shows off the air and then posting the result on the tracker. The time of posting and its comparison with the appearance of the same files or torrents on other fora or trackers is indicative of the kudos accompanying such acts: the faster uploader or the best quality of torrent documentation is the one that earns most respect from the rest of the community.
* Quality is also an important type of value produced. This may mean quality of digitisation or documentation or collection of digitisations that are then uploaded on the tracker. This is assessed both by the number and type of comments as well as by the ratings that users give to uploaders. It is important to note that the rating goes to the quality of the upload and not the actual content, which is not to be assessed by ratings but rather by the number of uploads.

3.1.4 Copyright status and other rights issues

* no copyrights are cleared
* the users create a great deal of meta-content for which there is no clear flow of rights or permissions. The value of this meta-content is effectively enjoyed by the community but is in the hands of the torrent tracker administrators that operate as custodians. This is in line with the technical nature of the medium that is fully decentralised with the indexing services hosted by the administrators and the users providing both content and meta-content.

3.8.5 Terms of access and use

* The content and meta-data are fully accessible and downloadable by all registered users. Registration opens at non-prespecified times in order to control the number and influx of users. The limited amount of users allows a more filtered participation and the smooth operation of the community.
* The user has to respect the netiquette of the forum and to share content in order to have a positive ratio (more than 1.0). This is done in order to increase the circulation of the material and decrease the phenomenon of hit and run or leeching or free riding of the common resource, that is the bandwidth and the actual content.

3.2 Case Two: BBC Creative Archive

3.2.1 Background

The BBC Creative Archive pilot ended in 2006 after temporarily releasing more than 500 pieces of digital content under Creative Archive Licence (CAL) draft scheme (Creative Archive Licence Group 2006). The project was set up in 2005 by the BBC, BFI (British Film Institute), Channel 4 and the Open University to make certain archive content available for the public to use under CAL.  Currently the BFI and the Open University are making archive film clips available for public use under this scheme.

3.2.2 Key content features

- Audiovisual content
-  Copyright and related rights

3.2.3 Value gains

- value from the creation of derivative works
- collective gains and cost reduction from not having to re-create the content.

3.2.4 Copyright status and other rights issues

- All rights have been cleared by the BBC
- Content with minimal copyright problems were primarily selected, such as factual documentaries where no music score has been used.

3.2.5 Terms of access and use

The five basic rules of CAL can be summarized as follows[6]:

A. Non-commercial

Anything you create using the available content must be for your own non-commercial use. This means that you can share it freely with family and friends and use the content for educational purposes. You may not, however, sell or profit financially in any way from the use of the content, for example, artists can't charge admission fees to exhibit work they've produced with the content.

B. Share-Alike

You are welcome to share the works (we call them 'Derivative Works') you produce with this content. If you do want to share your Derivative Works, please make sure you do so under the terms of the Creative Archive License, and make sure you 'credit' all creators and contributors whose content is included in the Derivative Works.

---

[6] http://www.bbc.co.uk/creativearchive/

C. Crediting (Attribution)

This is your chance to make sure everyone knows what you've done, but you also need to make sure that others who have contributed to a work (a Derivative Work) are credited too. It's up to you how creatively you acknowledge others' contributions!

D. No Endorsement and No derogatory use

We want you to get creative with the content we've made available for you but please don't use it for endorsement, campaigning, defamatory or derogatory purposes.

E. UK

The Creative Archive content is made available to internet users for use within the UK.

3.3 Case Three: Internet Archive

3.3.1 Background

The Internet Archive[7] is a non-profit organization that was founded in 1996 to build an Internet library. Its main initial goal was to offer permanent access for researchers, historians, scholars, people with disabilities, and the general public, to historical collections that exist in digital format. In late 1999, the organization started to grow to include more well-rounded collections.

3.3.2 Key content features

Now the Internet Archive includes <u>texts</u>, <u>audio</u>, <u>moving images</u>, and <u>software</u> as well as <u>archived web pages</u> in its collections, and provides specialized services for adaptive reading and information access for the blind and other persons with disabilities.

The content of the Collections comes from around the world and from many different sectors. It may contain information that might be deemed offensive, disturbing, pornographic, racist, sexist, bizarre, misleading, fraudulent, or otherwise objectionable.

3.3.3 Value gains

---

[7] http://www.archive.org

* The business model of the Internet Archive is based on donations.

* The main type of value produced is social value from providing access to a vast collection of material

* The internet archive also contributes to the preservation of the material

3.3.4 Copyright status and other rights issues

The Archive does not guarantee or warrant that the content available in the Collections is accurate, complete, non infringing, or legally accessible in user's jurisdiction.  The Archive makes no warranty of any kind, either express or implied. Though, the Internet Archive respects the intellectual property rights and other proprietary rights of others. The Internet Archive may, in appropriate circumstances and at its discretion, remove certain content or disable access to content that appears to infringe the copyright or other intellectual property rights of others.

3.3.5 Terms of access and use

* The Archive, at its sole discretion, may provide the user with a password to access certain Collections. The Internet Archive is committed to making its constantly growing collection of Web pages and other forms of digital content (the "Collections") freely  ("at no cost") available to researchers, historians, scholars, and others ("Researchers") for purposes of benefit to the public.

* A great part of the Internet Archive is made available to the end user under the six Creative Commons licences

3.4 Case Four: Broadcasting Archives (BBC)

3.4.1 Background

It is the main archive by the BBC part of which is made available online only to the UK users.

3.4.2 Key content features

BBC Archives[8] contain about 4 million items for television and radio. That is equivalent to 600,000 hours of television content and about 350,000 hours of radio. BBC Archives also have a New Media archive, which is keeping a record of the content on the BBC's websites, a large sheet-music collection, and commercial music collections. It also contains press cuttings going back 40 years and other kinds of items. BBC records and keeps everything for a minimum of five years. After this five years period all the news, the drama, the entertainment, the high value and expensive to make programs are kept following the BBC Archives selection policy.

---

[8] http://www.bbc.co.uk/archive/

3.4.3 Value gains

* preservation
* cultural goals are served through the dissemination of high quality and well curated content to the end user
* value returned to the licence fees payees

3.4.4 Copyright status and other rights issues

Due to rights restrictions some of the programmes in the BBC Archive Collections are only viewable from within the UK. The current agreement with copyrights holders allows the BBC Archive to stream programmes, so they can only be watched via the Archives' website. Finally, the Archives have a well organized policy to let a user know when a programme may include content unsuitable for children or when it may be harmful to view.

3.4.5 Terms of access and use

Users are not allowed to get free copies of programmes via the website, even if the programmes are already available to view on the website. If a programme has been broadcast within the last seven days, it may be available via BBC iPlayer. A special service, via the BBC Active, has been designed to fulfil the needs for academic and corporate training. BBC Active started as part of BBC Schools Radio in 1929. It was originally set up to produce simple teachers' notes to support the use of radio programmes in the classroom. It now publishes an extensive range of interactive resources based on BBC content, which support teaching and learning in primary and secondary schools, adult language learning and English language learning. In 2005 a joint venture was formed with Pearson to further develop these resources.

3.5 Case Five: British Library Archives

3.5.1 Background

The British Library (BL) is a non-departmental public body sponsored by the Department for Culture, Media and Sport of the United Kingdom. It is the national library of the United Kingdom and one of the largest libraries in the world.

3.5.2 Key content features

As a legal deposit library, it receives copies of all books produced in the United Kingdom and the Republic of Ireland. Its collection includes well over 150 million items, in most known languages. It receives 3 million new items every year. The British Library Collections consist of manuscripts, maps, newspapers, prints and drawings, music scores and patents. The Sound Archive keeps sound recordings from 19th century cylinders to CD, DVD and MD recordings. The Library's collections include around 14 million books along with substantial holdings of manuscripts and historical items dating back as far as 2000 BC. British Library operates the world's largest document delivery service providing millions of items a year to customers all over the world.[9]

3.5.3 Value gains

* value rests mainly with the actual content as well as its meta-data
* access to knowledge is also a great part of the value produced by the BL as well as part of its mission
* Using the website of the BL the users can query on:

  - 10.000 British Library web pages
  - 13 million records from the Integrated Catalogue
  - 90.000 pictures and sounds
  - 9 million articles from 20.000 top journals

3.5.4 Copyright status and other rights issues

The content (content being images, text, sound and video files, programs and scripts) of the BL website is copyright © The British Library Board. All rights expressly reserved. The users have to agree to abide by all copyright notices and restrictions attached to the content and not to remove or alter any such notice or restriction.

3.5.5 Terms of access and use

The content of the BL website can be accessed, printed and downloaded in an unaltered form (altered including being stretched, compressed, coloured or altered in any way so as to distort content from its original proportions or format) with copyright acknowledged, on a temporary basis for personal study that is not for a direct or indirect commercial use and any non-commercial use.

3.6 Case Six: BBC CenturyShare Project

3.6.1 Background

---

[9] Online catalogues, information and exhibitions can be found on BL website www.bl.uk.

The BBC CenturyShare project is jointly funded by JISC and the BBC Future Media and Technology (FMT), which is responsible for BBC's digital presence. The CenturyShare project is based on 'find, play and share', which is one of the BBC's Future Media and Technology strategies. The idea is to: (a) find BBC's content whether it is on or off the site; (b) play – or enjoy – it; and (c) share it to send it someone else, so that someone else finds it and the circle starts again. This project builds on the concept of liaising with different partners to produce products on the basis of the content that all collaborating organisations have, which is consistent with the key objectives of the SCA in promoting interoperability between and across different cultural sectors. For instance, instead of user-generated content the intention is to use the assets of the partners of the SCA, focused on specific themes, and gather them into one place to give people a way into the collections without going to the owners of them directly. The project is a proof of concept to determine whether it is a viable concept for SCA partners aiming to analyse, aggregate and augment cultural content. Ultimately, content will be displayed on a timeline, so part of the activity will be taking the material and seeing if there is a date description and then adding more to the description or more keywords etc.

The CenturyShare project is of particular interest as it operates in two layers: (a) it provides content collected from a network of providers; and (b) it allows the collection of meta-content created by the users.

3.6.2 Key content features

- Multiple types of content: images, video, audio, documents (literary works), diagrams (graphical works) and compilations of content

- Multiple sources of content under different licensing schemes

3.6.3 Value gains

- Allows users to identify public sector e-content that is most relevant to them

- Produces valuable metadata

- Links dispersed material along a timeline

- Increases e-content visibility and creates multiple access points

- Provides a platform for sponsors from the across the public sector to provide access to their content in one place

3.6.4 Copyright status and other rights issues

- Ownership of content will remain with the originating organisation of the content

- The responsibility for the clearance of content is managed by the participant organisations

- BBC acquires licences for the user-generated content

- Data-protection issues are thoroughly covered by the registration service agreement

3.6.5 Terms of access and use

BBC CenturyShare only provides a link to the e-content that is directly made available and licensed to the end-user by the organisation that owns the content.[10]

The metadata produced by the end-users are licensed to the BBC

3.7 Case Seven: British Library Archival Sound Recordings (BL ASR I and II) [11]

3.7.1 Background

The British Library's Archival Sound Recordings projects aim to digitise and make freely available 8,000 hours of digitised audio to the Higher and Further Education (HE/FE) communities of the UK. The projects are funded by JISC under its Digitisation programme. The core objectives of the project are to provide audio material for teaching, learning and research within various subject areas from history to ethnomusicology to science, across the broad range of HE/FE within a password-protected domain.

3.7.2 Key content features

- Multiple types of recordings: (a) unpublished recordings; (b) published commercial recordings; (c) oral history; (d) field recordings (sound scapes)

- Multiple types of works (published and unpublished) exist such as: (a) performances; (b) recorded literary works; (c) sound recordings; (d) musical works

---

[10] The BBC CenturyShare was a pilot that never managed to be fully implemented. Instead, the MemoryShare project is currently up and running that does not use content solely from the SCA partners. For an example of a memory (Amy Winehouse's death) see http://www.bbc.co.uk/dna/memoryshare/A86333330?s_fromSearch=ArticleSearch%3Fcontenttype%3D-1%26phrase%3D_memory%26show%3D8

[11] The British Library is one of the SCA sponsor organisations

- Multiple types of rights: (a) copyrights; (b) trademarks (on the brands of eg record companies); (c) personal data (eg in an oral history recording)

3.7.3 Value gains

- Educational and research value from making various forms of sound recordings freely available to the research community

- Cultural value from the preservation and dissemination of culturally important content that has not been previously published

- Increasing the visibility of the British Library archive and attracting a greater audience

- Allowing researchers to built upon primary material that is now made easily available

3.7.4 Rights ownership and obtained permissions

Rights are either owned by the British Library or effort is invested to obtain licences from the rights holders. The multiple layers of rights existing in each work often cause severe clearance problems and result in the emergence of a whole class of works without an identifiable owner (orphan works). More specifically:

- Clearance costs are high and unpredictable

- The clearance procedure affects the management of the whole project

- Clearance of rights is important not merely because of the legal liability risks but also in order to maintain the good reputation of the British Library

3.7.5 Terms of access and use

The content is made available to the public under two types of agreement, one for the general public and another specifically for HE/FE institutions.

- The material that is made available to the general public is licensed under a standard BL licence allowing end-users to copy the material for private, non-commercial and educational or research purposes. The licence does not permit adaptations or further dissemination of the work[12]

- The material that is made available to HE/FE institutions is licensed through the Archival Sound Recordings Sub-licence Agreement. Such a sub-licence allows under very specific conditions the copying and the limited distribution and adaptation of the content. More specifically:

---

[12] www.bl.uk/copyrightstatement.html

–        The circulation of the licensed content is allowed but only over a secure network, such as Athens, in the UK and between specific categories of users, as described in the sub-licence agreement. Authorised users are members of staff and students of the HE/FE institutions only

–        The sub-licence allows only educational and non-commercial uses of the licensed content

–        Authorised users, as defined in the sub-licence, are allowed to incorporate parts of the licensed content in their own work provided they properly attribute the right-owners and acknowledge the source

–        Public performance of the licensed content is only possible to the extent that the relevant additional licence has been provided by the relevant collecting society

3.8 Case Eight: National Library for Health eLearning Object Repository (NLH LOR)[13]

3.8.1 Background
The National Library of Health (NLH) eLearning Object Repository (LOR) project is part of the National Health Service (NHS) Institute for Innovation and Improvement. Its main objective is to provide access to standards-based e-learning objects via a cross-searchable and browseable open web interface. All registered members of the NHS workforce will be able to search the repository and download objects that are on Open Access for use within local Learning Management Systems (LMS).

3.8.2 Key content features
- Multiple types of content: images, video, audio, documents (literary works), diagrams (graphical works) and compilations of content
- Multiple sources of content provided under different licensing schemes

3.8.3 Value gains

- To improve the search and identification of content on the platform

- To reduce the duplication of effort in the production of learning objects/content by the participating

organisations/communities

- To share educational material

---

[13] NLH LOR is one of the SCA sponsor organisations

- To facilitate the improvement of existing material

- To link together different types of material

- The core value of the NLH LOR project comes from reducing redundancy in the production of content and from 'recycling' resources from various communities. As a result, the value of the project increases in proportion to the ability to identify, share and repurpose the content stored in the repository

3.8.4 Rights ownership and obtained permissions

- The copyright in the NLH website belongs to the NHS institute for Innovation and Improvement[14] unless stated otherwise[15]

- The content uploaded by users of the NLH LOR is not licensed specifically to the NHS but, instead, it is directly licensed to the end-user through one of the three Creative Commons Licences made available through the website[16]

- The contributor of the material is responsible for IPR clearance

3.8.5 Terms of use and access

- Three Creative Commons (CC) licences, all containing the Non-Commercial licence element, are the ones used for the dissemination of the content:

– Creative Commons Attribution Non Commercial (CC–BY–NC):[17] this is a non-exclusive licence allowing the licensee to copy, distribute, transmit and adapt the original work under the condition that the work is: (a) attributed in the manner specified by the author of the work or the licensor and in accordance to the terms of the licence; and (b) it is not used for any commercial purposes

– Creative Commons Attribution Share Alike Non Commercial (CC–BY–NC–SA):[18] this is a non- exclusive licence allowing the licensee to copy, distribute, transmit and adapt the original work under the conditions: (a) that

---

[14] www.institute.nhs.uk/index.php

[15] See for example statement in www.library.nhs.uk/mylibrary/default.aspx

[16] www.creativecommons.org

[17] http://creativecommons.org/licenses/by-nc/2.0/uk/legalcode

[18] http://creativecommons.org/licenses/by-nc-sa/2.0/uk/legalcode

no commercial use of the work is made; and (b) that the work is attributed in the manner specified by the author of the work or the licensor and in accordance to the terms of the licence. The licensee is also allowed to build upon[19] the original work, provided they share the resulting work under the same conditions

– Creative Commons Attribution Non Commercial No Derivatives (CC–BY–NC–ND):[20] This non- exclusive licence allows the licensee to copy, distribute and transmit the work under the following conditions: (a) the work is attributed in the manner specified by the author of the work or the licensor and in accordance to the terms of the licence; (b) the work is not used for commercial purposes; and (c) the licensee does not alter, transform or build upon the work. This is the most restrictive for the licensee Creative Commons Licence as it confers the most limited set of permissions to the licensee

☒ The non-commercial element was chosen as one expressing the non-commercial nature of the project

☒ The employment contracts defining the ways in which NHS employees may use material on the NLH LOR may be in conflict with the CC licences[21]

### 3.9 Case Nine: Great Britain Historical Geographic Information System/ Vision of Britain Through Time

#### 3.9.1 Background
The Great Britain Historical GIS (or GBHGIS) is a spatially-enabled database that documents and visualises the changing human geography of Great Britain, mainly over the 200 years since the first census in 1801. The project is currently based at the University of Portsmouth, and is the provider of the Vision of Britain through Time (VoB) website.[22] The project is involved in the digitisation of a wide range of geographic and demographic data that are included in the GBHGIS.[23] The objective of the project is to make the data available to the widest possible range of users through a variety of channels and encourage their reuse in different contexts. For instance, the digitised and compiled data may be either downloaded from UKDA (the UK Data Archive)

---

[19] Eg extend, reuse, repurpose

[20] http://creativecommons.org/licenses/by-nc-nd/2.0/uk/legalcode

[21] The Creative Commons licences do not set any limitations where the content is to be used (eg within or outside the NLH network)

[22] http://en.wikipedia.org/wiki/Great_Britain_Historical_GIS

[23] The Joint Information Systems Committee (JISC) funded early development work on the GBHGIS web-based mapping tools, under JTAP project
JTAP 1/320, 'Authoring methods for electronic atlases of change and the past', and contributed to boundary mapping and data entry.

[24]and EDINA's (Edinburgh Data and Information Access)[25] UKBORDERS (United Kingdom Boundary Outline and Reference Database for Education and Research Study)[26] service or may be viewed on the Vision of Britain website.

3.1.2 Key content features
*   Data intensive content (data and data compilations)
*   Maps and graphics
*   Material from the 19th and 20th centuries (material in the public domain)

3.1.3 Key value gains
*   Through the VoB service the visibility and usability of data, especially for non-expert users, is increased
*   By allowing the downloading of data in raw form (through UKDA and EDINA UKBORDERS), it is possible to link them with other related services (eg archives, other GIS services) and thus achieve their maximum utilisation
*   Different channels of making the data available serve educational and research objectives
*   As the access to the data becomes easier, added cultural and historical value is provided to non-
professionals (eg amateur local historians, lay users)
*   The availability of data in different forms could potentially create a market for individuals interested in family and local history or location-sensitive services

3.1.4 Rights ownership and obtained permissions
*   Most of the works used for the project are currently out of copyright, although some of the works will be protected by Crown Copyright
*   There are a variety of copyright owners within the VoB project. These include:
–   The copyright ownership of Census data from 1961 to 2001 belongs to National Statistics, for England and Wales, and to the General Register Office, for Scotland. These agencies also supplied the VoBs with detailed maps of modern census reporting areas
–   The copyright in some of the historical photographs used within the VoB belongs to English Heritage
–   The copyright in the text interpreting statistical themes belongs to Humphrey Southall 2003, 2004
–   The copyright in the maps created by the Land Utilisation Survey of Great Britain belongs to L. Dudley Stamp/Geographical Publications Ltd, while the scanned images of these maps, for England and Wales, to the Environment Agency/Defra, and for Scotland to the Great Britain Historical GIS
*   The data used for the project have been collected for a period of about ten years. In this period, the data collection and compilation have been funded

---

[24] www.data-archive.ac.uk

[25] http://edina.ac.uk

[26] http://edina.ac.uk/ukborders/description

by a variety of projects and the individuals collecting and compiling the data have been employed by different academic institutions. As a result, there are potentially a number of rights holders for the data
* Issues of institutional ownership and transfer of rights have been resolved in the following ways:
–	By ensuring that the Principal Investigator,[27] i.e. the person heading the research project, obtains a licence from the academic researchers who hold copyright in the transcriptions
–	By assigning or licensing all copyright to an organisation[28] that exists irrespective of any project transformations

### 3.1.5 Terms of access and use

The content found on the VoB website is not licensed to the end-user under a specific licensing scheme. It only contains detailed copyright notices regarding each of its components.[29] Consequently, the use of the content is governed by the rules of fair dealing as defined in the relevant legislation, ie content can be used for non-commercial research or private study.[30]

The content made available through the UKDA and EDINA BORDERS services is licensed under the Census End User Licence (EUL).[31] The key terms of this licence agreement are as follows:
*	Data can only be used for personal, research, educational and non-commercial purposes
*	Registration is a requirement for using the content
*	The data cannot be further disseminated
*	Personal information must be kept confidential
*	Attribution and acknowledgement is made in accordance with the terms and conditions of the licence

## 4. Models of permission and content flows

### 4.1 General observations

---

[27] Professor Humphrey Southall

[28] According to the VoB website (www.visionofbritain.org.uk/footer/doc_text_for_title.jsp?topic=credits&seq=4) 'The resource as a whole is © Great Britain Historical GIS Project 2004', the GBH GIS being a network of collaborating academic researchers.

[29] www.visionofbritain.org.uk/footer/doc_text_for_title.jsp?topic=credits&seq=4

[30] Sections 29 and 30 of the 1988 Copyright Designs and Patents Act

[31] www.data-archive.ac.uk/aandp/access/licence.asp

Different IPR management approaches appearing in the projects examined in the seven case studies may be abstracted in three main models of works and permission flows.[32]

Flows of permissions related to moral rights do not appear in the diagrams. This is because in all cases examined in this report moral rights remain with the creator of the content.

Three models of content and permissions flows are presented in this section. Each model is named after the key characteristic of the way in which the flows are structured. The three models are as follows:

The 'Star-Shaped' model
The 'Snow-Flake' model
The 'Clean Hands' model

Such models are illustrative of the ways in which IPR management may enable or hinder the flow of e-content. They also constitute a basic typology of the ways in which different models of IPR management could facilitate different types of value production. Finally, each model may be associated with different organisational objectives. In that sense, such models could inform the way in which IPR policy and strategy are formed.

There is no one-to-one correspondence between models and projects. For example, in each project more than one model may appear and one flow model may be used in more than one project.

4.2 The 'Star-Shaped' model
The Star-Shaped model may be applied to collections and dissemination of permissions and content.

4.2.1 Collection of content and permissions
The star-shaped model involves a central entity that is responsible for the acquisition of the content and the required licences from the content providers and/or other rights holders, both of whom may be individuals, organisations or other projects.

The central entity that resides at the centre of the star is the one responsible both for the clearance of the rights and the curation of the material. The flows of permissions and works follow the same direction, although they can follow different paths, ie flowing from the supplier to the central entity. This is because it is likely that the rights owner and the content provider may be different, and the supply of each may be made at different times, particularly when rights are cleared for legacy material already owned by the central entity. This means that the acquisition of permissions may follow a push or pull model, ie either the central entity is in possession of the content and asks the relevant permissions from the rights holder or the rights holder deposits the material with the central entity agreeing to license the work under specific terms and conditions set by the central entity (see Diagram I).

---

[32] We use the term 'permission flows' to denote flows of copyright licences between different users and stakeholders in each of the models. A flow does not necessarily mean that the licensor is stripped of all their copyrights. In most cases, the copyright owner only awards a licence, ie a set of permissions, that flows within the boundaries of the project. The exact terms and types of licences are presented in greater detail in the appendices of this report. The concept of permission mainly refers to licences, but it is broader than mere licensing. For example, in the case of the NEN Repurpose project, permissions are sought from the parents for the use of the works of their children.

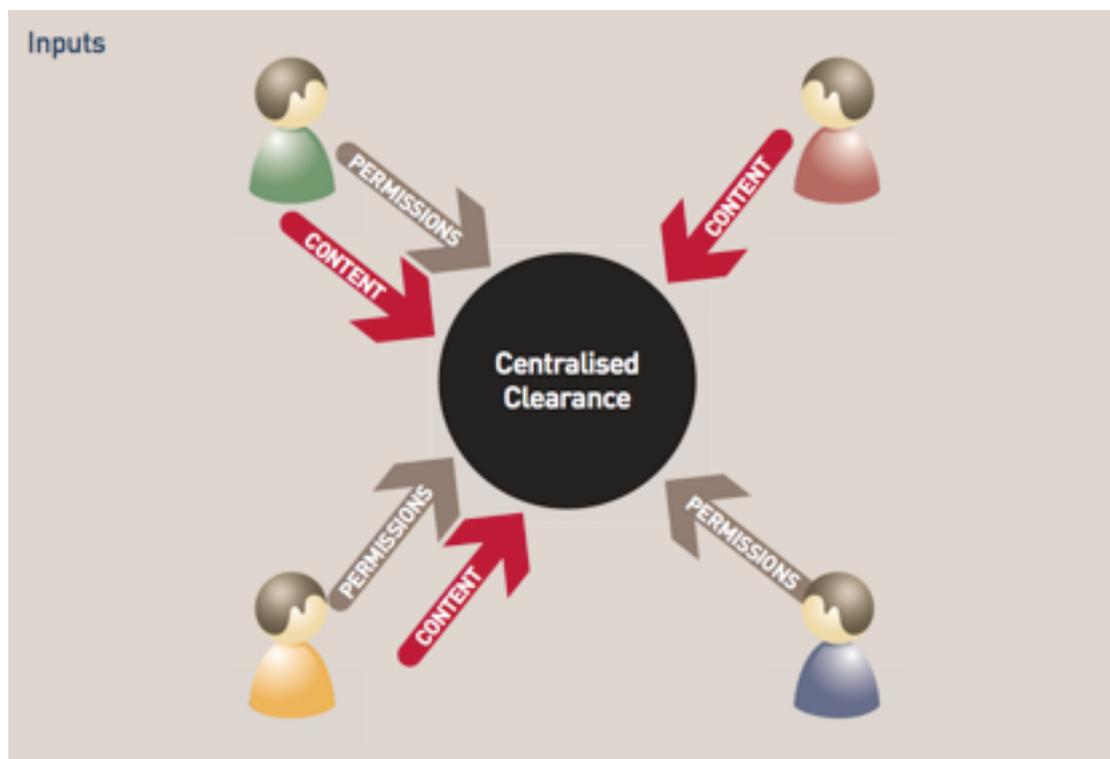

Diagram I

4.2.2 Impact

Most projects involving digitisation of analogue material, particularly in the context of museums and archives, are organised using the star-shaped model.

The star-shaped model reduces risks from copyright infringement as the process of copyright clearance is managed at a single point. At the same time, the cost for the organisation managing the process increases, as such a model requires a specialised service or unit to perform the function. As a result, this is a model that could be beneficial for a large organisation that can achieve economies of scale, but may not be sustainable for small and medium size organisations. In the latter case, a star-shaped model may lead the organisation to a strategy of avoiding digitisation of works that require any copyright clearance in order to reduce costs.

For an organisation to be able to benefit from such a model, it is necessary to establish standardised clearance processes and risk management protocols, such as those developed as part of the SCA IPR Toolkit. Such strategy will allow the organisation to accrue knowledge from the accumulated clearance experience. It is necessary to properly document the clearance process so that there are records of the material cleared. Ideally, the metadata from the rights documentation should be in a standard form so that other institutions or projects can make use of them.

For small and medium size organisations it is necessary to port ready-made clearance and risk- management procedures and customise them to their personnel and technology requirements. Another solution would be to establish a clearance service for a specific sector (eg museums) at a national level and thus reduce the costs for the individual organisations.

4.2.3 Example

The star-shaped model may be applicable even in cases where the organisation collecting the content and the permissions keeps transforming. This is the case of the VoB, where the organisation performing the collection has changed several times due to transformations in the project (see diagram II).

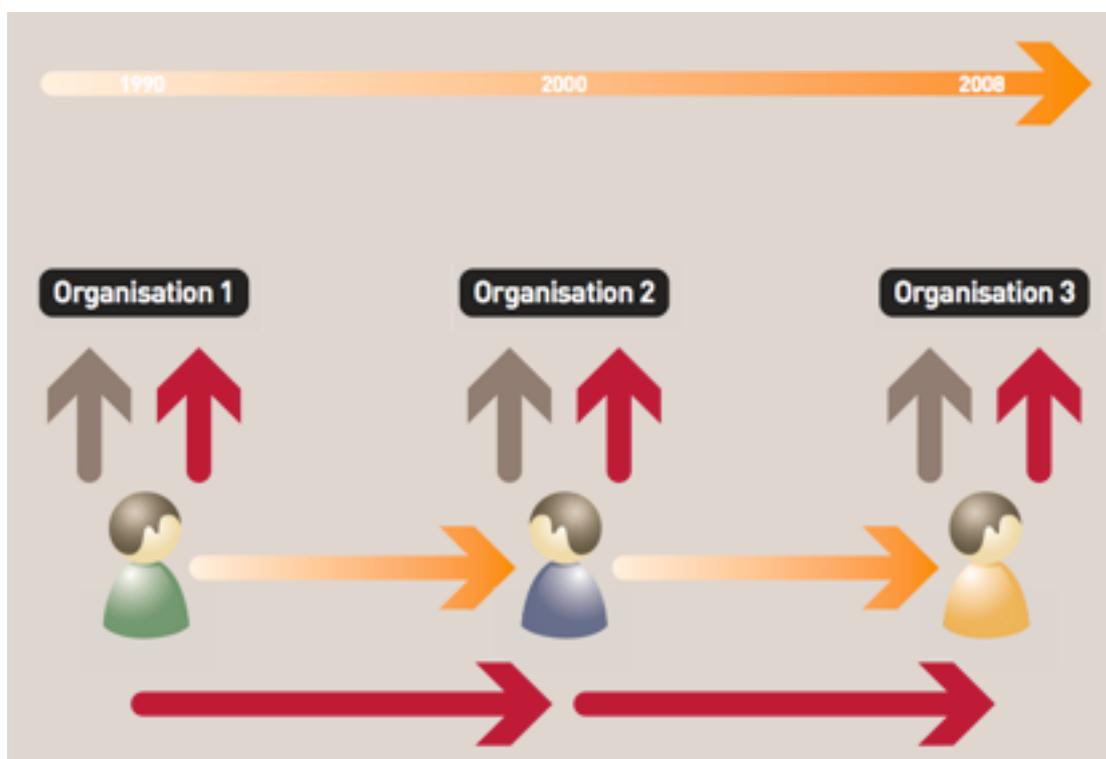

Diagram II

In this case, the continuity of the VoB project has been preserved by ensuring that a single point was responsible for the collection of content and permissions that the star-shaped model provides. This point of collection functions de facto as a rights repository and constitutes a solution for ensuring the permissions and content have been collected and the project may continue to exist (see diagram III).

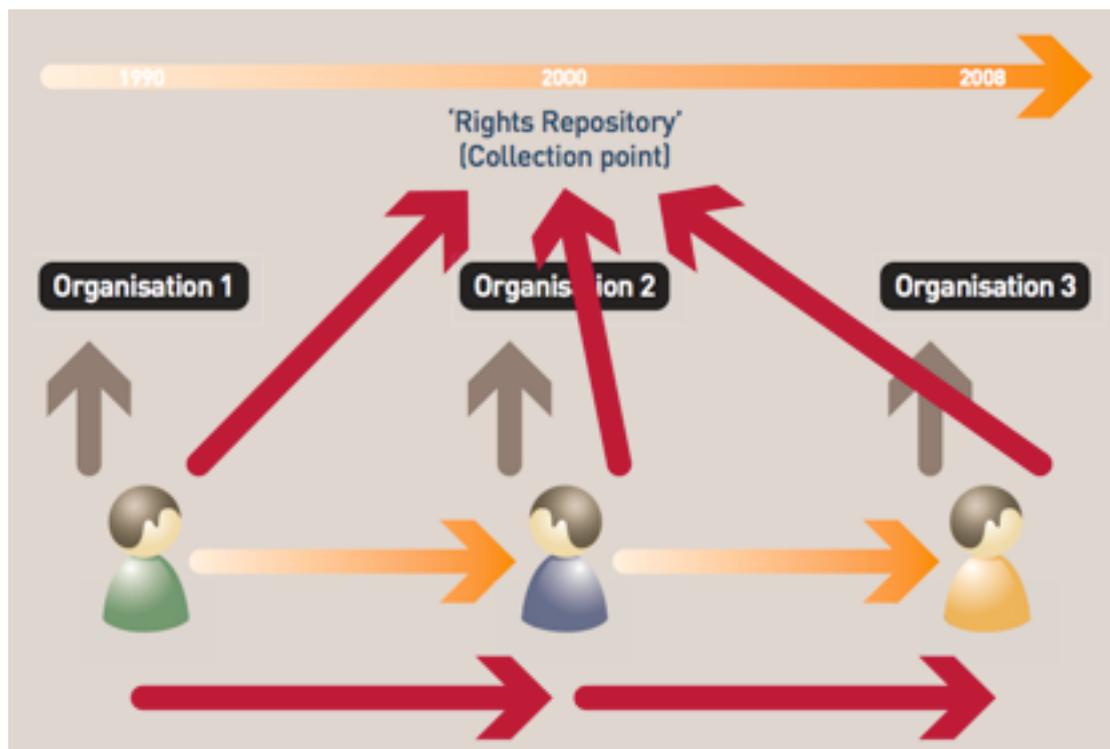
Diagram III

4.2.4 Dissemination
Digitisation projects
* Document and standardise clearance processes
* Put a risk assessment and management scheme in place
* Standardise metadata to facilitate communication between different institutions
* Establish a clearance service per sector (eg Museums) or region in order to achieve economies of scale.
The dissemination of content may also fit under the star-shaped model. In such cases, both the distribution and licensing of content is managed by a single central organisation. In this case, there are three broad scenarios of content and licence distribution under the star-shaped model:

- Public internet distribution
- Walled garden distribution, ie restricted distribution
- Hybrid public/walled garden distribution

4.2.4.1 Dissemination over the public internet
When content is made available over the internet the following are most common characteristics of its dissemination (see diagram IV):
*	There is always some form of licence specifying the permissible uses
*	The End-User Licence Agreements (EULAs) are custom-made licences that reflect the policy and strategy of the specific organisation

\*   The EULAs allow only private and non-commercial or educational uses. No super-distribution, ie further dissemination by the user or publishing on their private website is permitted. Repurposing is usually prohibited as well
\*   The quality of the digital surrogates is normally of low quality. For instance, low resolution images or videos, low bit-rate sound recordings

\*   In cases of audio or video, the content is only made available for streaming, not downloading
\*   No Technical Protection Measures (TPM) are used for still images or audio (Akester 2006). However, some of the audiovisual content is protected with TPM and downloading may be allowed only for a limited amount of time (eg BBC iPlayer)
\*   As a result, the content, both technically and legally, cannot to be repurposed either by end-users or other public-sector organisations

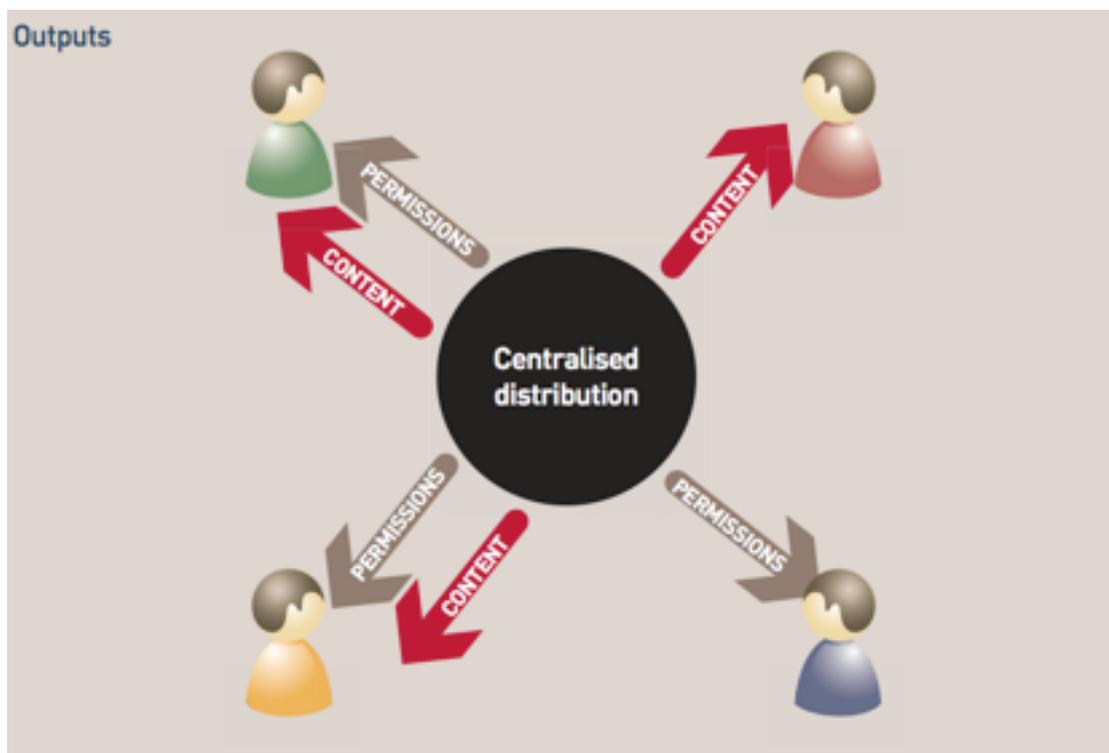

Diagram IV

4.2.4.2 Walled garden distribution
When content is made available over a controlled/secure network the following are the most common characteristics of its dissemination (see Diagram V):
\*   The dissemination of content over a secured environment is expressed in the related EULAs and the technologies of distribution. The EULAs are custom-made licences that reflect the funding conditions of the specific digitisation programme (eg the BL SA I was only made available to FE/HE students) or the charter of the digitising organisation (eg BBC content is normally made available only within the UK). The technology normally allows access to the content either through a specific gateway or on the basis of the IP address. For example, in the case of the BL ASR I project, the digital audio

recordings are made available only to UK HE/FE students and members of staff through the Shibboleth service; the BBC audiovisual content is only made available to users having a UK Internet Protocol address

*   The rights awarded to the users are normally greater than those found over the public internet. They normally include rights of reuse within the specific network. Such is the case of the BL SA II project, where the content is made available for reuse only within the secure network. Such an approach may be problematic as it creates pools of content that because of the licensing terms may not be legally interoperable with content that is reusable under a standard public licence, such as the Creative Commons licences

*   No technical protection measures are used on the actual content but access is allowed only to authorised users over secure networks

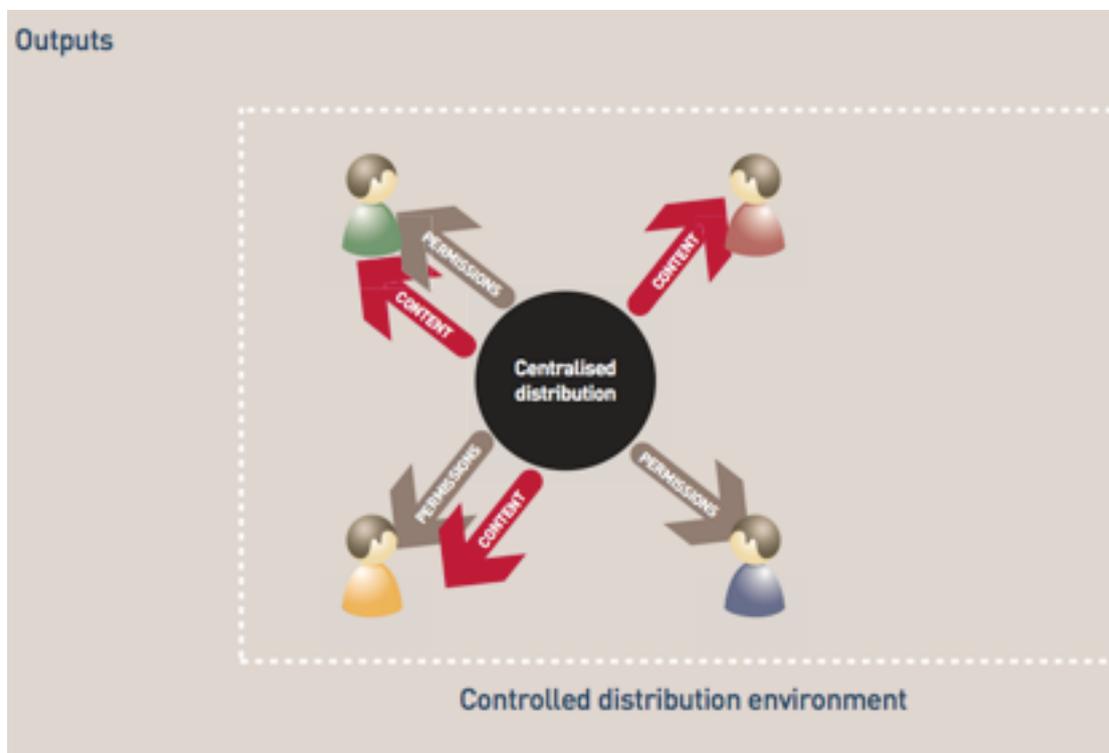

Diagram V

4.2.4.3 Hybrid public internet/walled garden distribution

This is the case when content is made available by the same central point both to the public internet and over a secure network (see Diagram VI). The case applicable in this model is the BL ASR II project. In such a scenario:

*   Different sets of content are distributed over public and secure networks, with premium or full content being provided over the latter

*   Different sets of rights awarded to the two types of users (public/within the walled garden). In the case that reuse rights are granted to users within the walled garden, the 'licence dilemma' appears

*   If a standard public licence allowing reuse is used (eg the Creative Commons licences), then the content may be legally and freely disseminated and reused on the public internet

*   If a custom-made licence allowing reusability is employed, then it will be very complex legally (and subsequently very expensive) to combine the

walled garden content with free internet content. The creation of content islands may be desirable in the short term but may cause substantial clearance problems or may even make the recombination of the content unusable in the long run

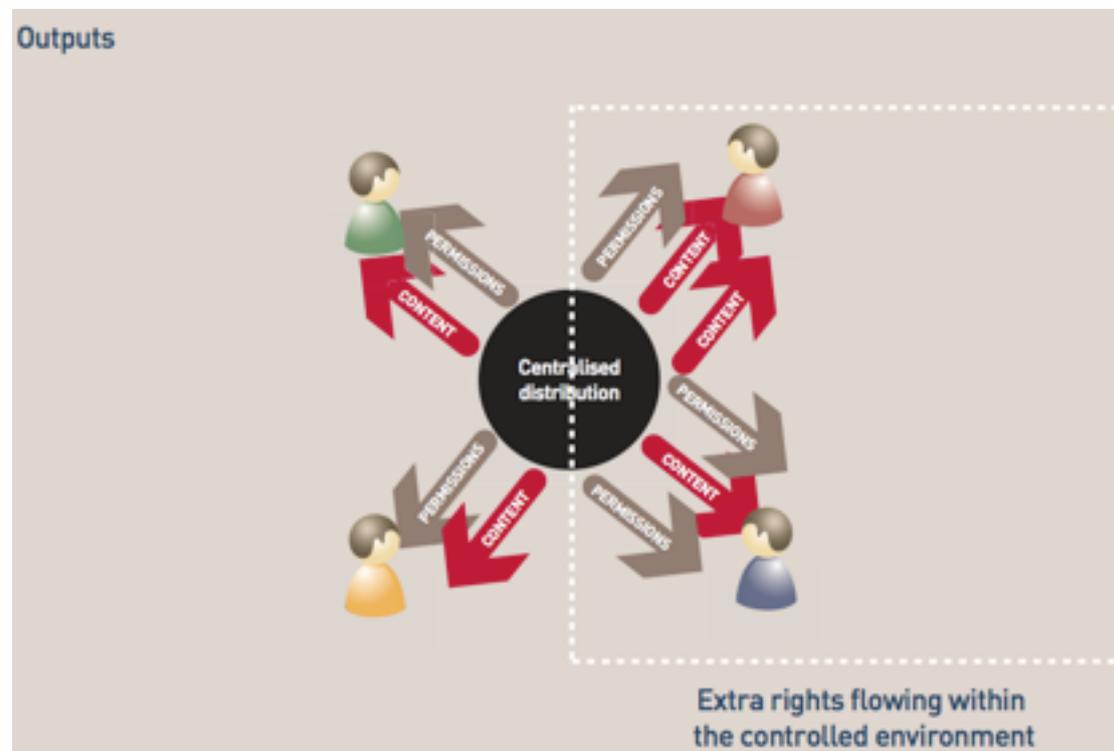

Diagram VI

4.3 The 'Snow-Flake' model

In the snow-flake model (diagram VII) the clearance of rights (obtaining permissions) and acquisition of content is organised in clusters: rights are cleared and content is aggregated first locally, then in clusters of local units and finally in a central hub. This type of collection appears in the NLH project and in some sense in the BlueGreece projects. It is a model that allows the reduction of clearance costs for the central organisation: the costs of clearance are primarily covered by the local organisations or at the cluster level. The central organisation oversees and manages the whole process but is not involved in any clearance itself.
Standardised risk management and clearance procedures are quintessential for the success of this model. The central organisation needs to have in place such procedures in order to ensure that the risk of copyright infringements is mitigated.
The snow-flake model is particularly popular in projects that:

* Are geographically dispersed
* Have multiple units
* Deal with more than one type of rights (eg copyright, personal data, protection of minors etc) that can be acquired and managed locally

4.3.1 Example
The snow-flake model is primarily used for content aggregation and rights clearance and does not have to be also followed in the distribution and licensing of the content. The latter may follow a hybrid snow-flake and clean hands model, as is the case of the BL SA II project. In this project, once clearance is completed in the local level:

* The content is licensed to the central entity
* There is cross-licensing of the content between the consortium parties
* Each consortium party decides by itself how to further license the content

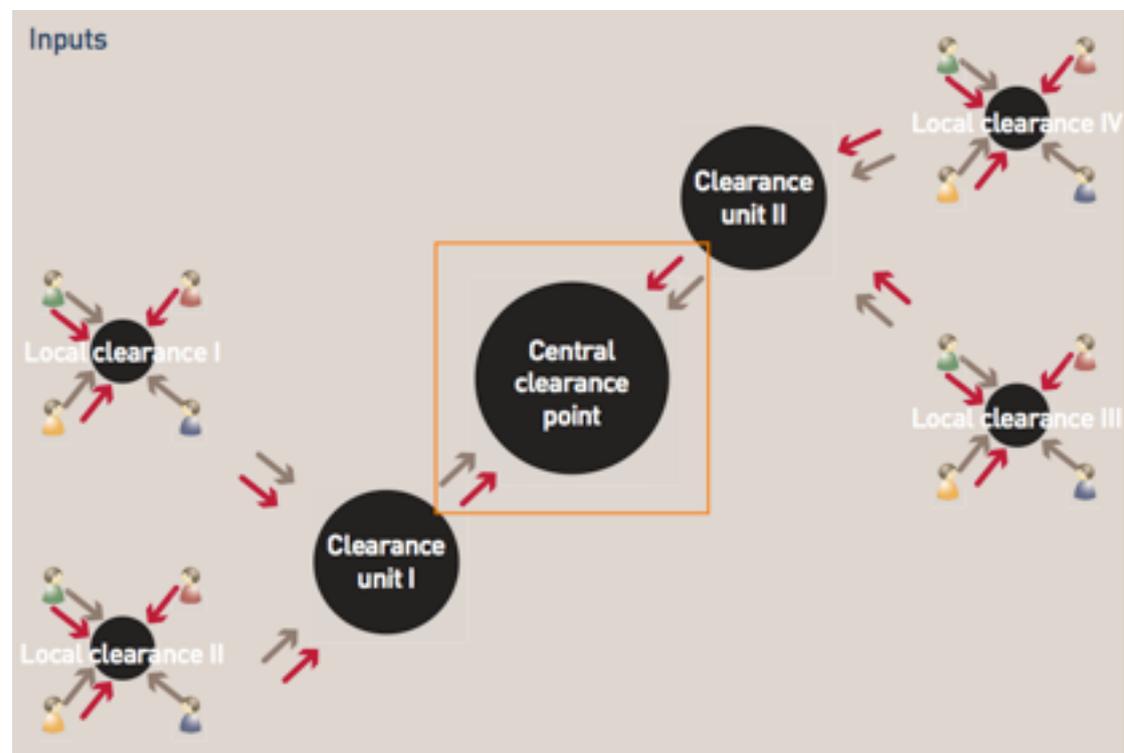
Diagram VII

4.4 The 'Clean-Hands' model
This is the model where the flows of rights and content follow entirely different paths. The content is normally collected and may be downloaded from a single point, whereas the licences flow directly between the users. The central organisation does not deal with copyright at all and that is why we use the metaphor of clean hands to describe the model (see diagram VIII).

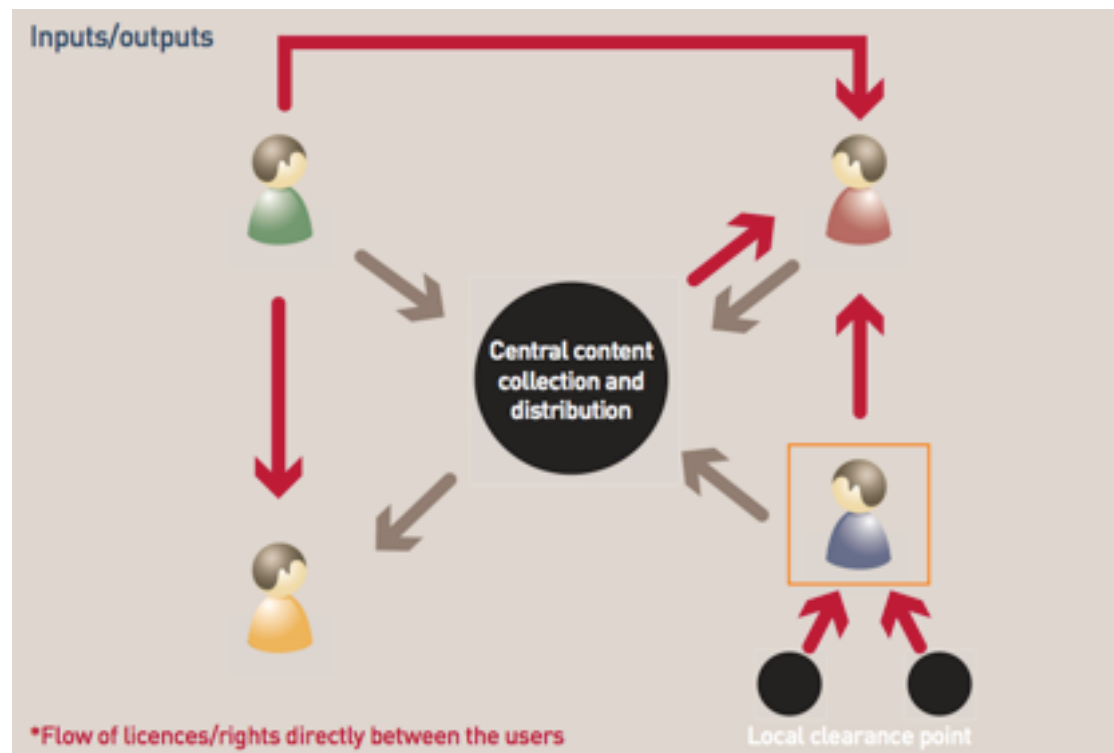

Diagram VIII
The key characteristics of the clean hands model are as follows:

\*      The clean hands model is not necessarily concerned with the aggregation of content or licences but rather with facilitating the respective flows (of content and licences between the users). The aggregation of content could take place in a centralised fashion and hosted by the central organisation (eg in the case of the BlueGreece, Creative Archive and NLH LOR projects), or to be directly managed by the participants of the system (eg CenturyShare project). The central organisation is not at all concerned with acquiring any licences over the content. In this model the central organisation only ensures that the end-users have the necessary permissions supplied by the rights owners
\*      The clearance of the content is pushed at the ends of the network or on the contributors of the content. These may be either individuals, legal persons or other projects. They are responsible not only for the copyright clearance but also for obtaining any other required permission such as Prior Informed Consent or personal data clearances
\*      The main risk management approach followed by the central organisation relies on their lack of direct involvement in obtaining any permissions for themselves and clearly stating in the service registration agreement that the end-user is responsible for the clearance of rights. Additional necessary measures include the provision of proper disclaimer clauses and clear notice and take-down procedures

4.4.1 Impact
\*      This particular model can result in the possibility of the 'licence pollution' phenomenon. Specifically, in a reuse scenario the copyright licences used

have to be compatible with each other, otherwise they will lead to derivative works infringing the copyright of the content on which they are based. For example, all Creative Commons licences are not compatible with each other and if they are used in a service (eg in the NLH LOR project) it is necessary that some minimum care is taken to inform the users accordingly. This may be done by ensuring that in the case of uploading a derivative work, the user is obliged to name the content sources and their respective licence. The system then should automatically inform the user about the compatibility of the source licences

\*	In any reuse scenario, the rights information should refer to the work, not the creator (see diagram IX). Hence, it is necessary to have metadata attached to each work making explicit:
–	Which works it is based on
–	In which works it has been used

\*	Overall, it is advisable to use standard licences and metadata so that linking with other organisations
and projects is possible
\*	The more rights are offered to the licensee, the more the need for:
–	Attribution
–	Provenance
–	Quality assurance
–	Adherence to data protection rules, processes for protecting minors and Prior Informed Consent rules

4.4.2 Examples
The clean-hands model is adopted in the following cases:

The central organisation is interested only in aggregating content from various other organisations or projects that provide content under a variety of licences. In this case, the central organisation may not even host the actual content: it may only provide the links to the content and perform the functions of aggregation and curation. The value, in this case, derives from increasing visibility and associating content with other related content. Therefore any metadata created are normally owned by the central organisation. This is the case of the CenturyShare project
The central organisation is interested in the reuse of content provided either by end-users, other projects or organisations. The value comes from the reuse and incremental improvement of content. These are the cases of the Creative Archive and BlueGreece projects.

\*	The central organisation hosts only user-generated content that freely flows on the internet. Value derives again from building on existing material and collective development. By pushing the rights clearance at the ends of the network the organisation decreases clearance costs and mitigates risks. It is not responsible for managing the complex ownership questions that are likely to appear. In this case standardised licences, such as the Creative Commons licences, are used. The most relevant related projects are the NLH LOR project.

4.4.3 Value
The main sources of value in the clean hands model are:
*   The cultivation of communities
*   The production of metadata
*   The linking of relevant content
*   Reduction of redundancies
*   Incremental innovation

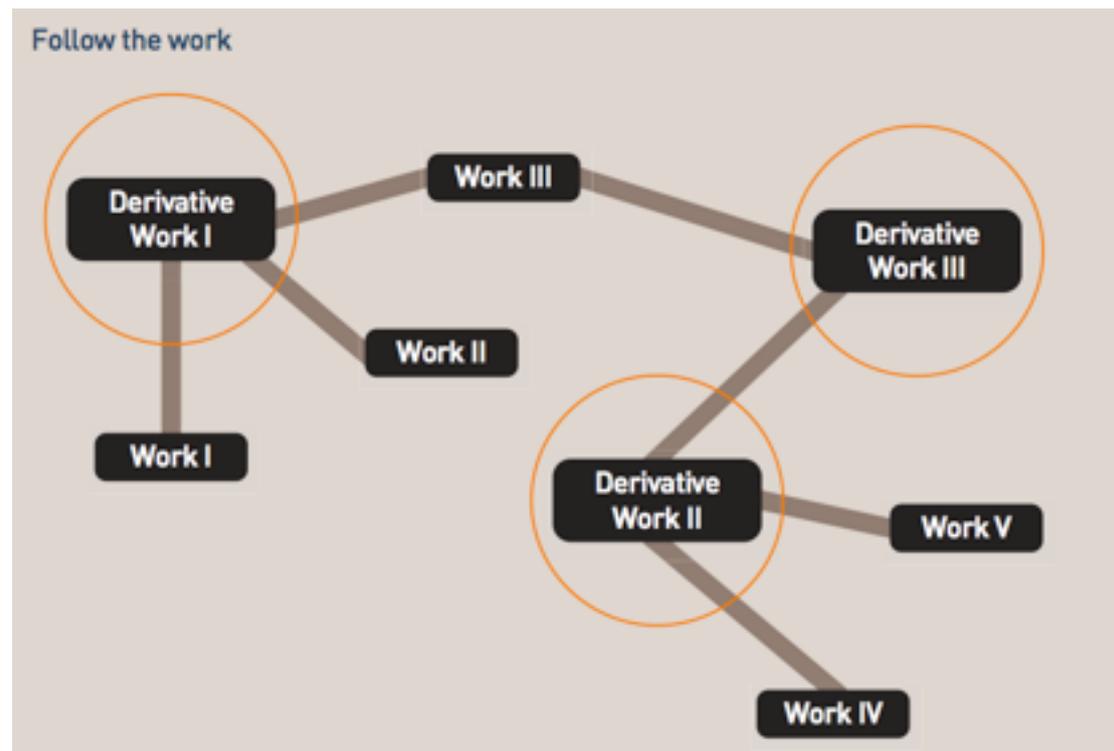

DiagramIX

4.5 Conclusion
Irrespective of which model IPR model is to be followed, a suitable copyright management framework needs to be implemented to ensure that basic procedures and decision-making rules can be widely adopted. This will ensure that staff and users understand the nature of the permissions that are being granted regarding access and use of content.

**5. Key Findings and conclusions**

5.1 Key value types identified
There is a variety of different value types identified in the case studies. The following list covers the main value types that are likely to be encountered.

5.1.1 Types of non-monetary value
*   Cultural dissemination and preservation
*   Educational
*   Reputational

* Quality
* Audience creation
* Relevance of material
* Collective memory
* Sustainability

5.1.2 Types of monetary value
This is a value associated with revenue, sustainability of the project and the ability of being able to secure future funding.
* All projects considered monetary value NOT as the key value to be achieved but rather as something that may be either useful in the future or necessary for sustainability purposes
* The production of monetary value appeared as a consideration in the form of ensuring that existing funding will continue and new public funding will be provided. As a result of the source of the monetary value being of public nature, the key objectives of all such projects has been to achieve public-serving purposes. Such purposes almost invariably require increasing access and allowing reuse of content.
* Finally, it means that monetary value and content or rights are not directly exchangeable. For instance, the Creative Archive is funded by public money and the Internet Archive through donations in order to make content freely available for sharing and repurposing. The users of such services do not directly pay for their use.

There are various perceptions of value types in different levels of hierarchy within the same organisation and are greatly contingent upon risk perceptions. For instance, middle management in a museum may consider provision of access in all possible ways the key objective, whereas the members of the governing trust may consider the reputation of the institution and the collection of material as the primary objective. Also the perception of value and risk greatly differ between the copyright specialists within the organisations and the rest of the staff interviewed in this study.

5.1.3 Conclusions
Although the value type identified from the case studies is not necessary monetary, there are inevitably costs in the production and dissemination of content that have to be somehow covered. These costs involve rights clearance costs (tracing rights holders, paying copyright fees for the acquisition of licences) and personnel costs (eg for the curation of the aggregated content or the monitoring of the service).

Even when the value produced is recognised as monetary, other forms of value, such as cultural and educational value, are equally important for the success of the project.

5.2 Funding and IPR management
Funding plays a key role in the formation of the project's IPR policy. It may define the broader framework of managing IPR or require the licensing of the content to the funding organisation (eg the BBC Archives are made available only to UK citizens and the Internet Archive makes all its content freely available as they have different funding mandates). Overall (Pollock, Newbery et al. 2008):

\*   Funding contracts could be used as a way to ensure licensing compatibility among different organisations and facilitate the cultivation of a common information environment
\*   Clauses requiring licensing to the organisation providing the funding need to be thoroughly re-assessed in order to ensure that they cover only the material for which clearance has been secured
\*   The problem of IPR clearance has to be addressed in the level of funding contracts in terms of:
–   Ensuring that clearance of rights is also funded, sometimes even as an auxiliary project
–   Acknowledging the time management implication that any clearance procedure entails
–   Funding training programmes for the staff in the areas of general IPR understanding, copyright, Open Licensing, Data Protection, confidentiality and prior informed consent agreements

5.2.1 Conclusion
Funding initiatives should take into consideration the costs and time management implications of clearance procedures and the need for training of staff on IPR management and other rights (eg personal data) issues. Such issues are outlined within the SCA IPR Toolkit.

5.3 Risk management
Risk management strategies do not exist in all projects. The existence of a comprehensive risk strategy is mainly contingent upon two factors (Hutter 2006):
\*   The experience of risk management in the organisation where the project is positioned: the more experienced the organisation, the more likely is that the specific project will also have a risk mitigation strategy in place. For example, BBC and NLH have comprehensive risk management strategies in place and this is found also in the CenturyShare and NLH LOR projects
\*   The degree to which the project involves acquisition of licences by the organisation managing the project: the more licences the organisation managing the project acquires, the more likely it is that a risk mitigation strategy will be in place. For example, the BL SA I and II projects have a very comprehensive risk management tool in place as it acquires rights, whereas the Internet Archive is the opposite case as the rights are transacted directly between the creator and the end-user with the project only providing some basic infrastructure. Finally, the GreekBlue project seems takes a very liberal approach in the sense of not interfering wit the torrents the users are sharing and only blocking very recent movies (especially porn) as this seem to be the riskiest types of content to be uploaded without prior permission from the copyright holders.

5.3.1 Summary

\*       Risk management strategies need to operate at the level of individual rights (eg right of reproduction, right of attribution) (Ciborra 2004)
\*       Dates of expiration of rights should always be recorded
\*       The permissions acquired by the organisation should be equal or more than the permissions the organisation grants to the user of its services
\*       Risk management strategies need to be developed in the form of toolkits made available to different organisations to adjust them to their own projects (such as is the case with the various SCA toolkits) (Lezaun and Soneryd 2006)
\*       Risk management strategies need to be evaluated in conjunction with the intended value production streams
\*       Training in IPR risk management processes have to be developed with respect to (Taylor-Gooby and Zinn 2006):
–       Staff of organisations managing IPR-related projects
–       Users of services that require them to do some form of pre-clearance or clearance of material
–       Project partners involved

### 5.3.2 Conclusion

Risk management approaches need to be developed in the form of ready-made toolkits, and risk management training is required not only for the staff of organisations managing IPR but also to users performing clearance procedures. The SCA IPR toolkit addresses such concerns.

### 5.4 Content and rights identification

Works and rights identification is a necessary step toward the development of risk management approaches. It is the stage for example, at which the extent of the orphan-work problem may be identified and therefore measures implemented to manage risk.

The existence of multiple layers of works and rights in the same object has increased the costs of clearance of rights because the number of authors to be identified and the rights to be negotiated has increased. The more layers of works/rights an object contains, the more likely it is that no value, monetary or not, can be created. This is a phenomenon appearing particularly in the context of digitisation projects such as the BBC Archives and the BL ASR projects. This phenomenon is a direct result of the clearance costs for content comprising of multiple types of rights. In projects like BlueGreece, the organisation managing the project does not have the resources to complete the clearance for such works, whereas in projects like BL ASR, the time limitations that the project management imposes make the clearance of such content very problematic. For instance, a sound recording with performance rights, sound recording rights, literary works and musical works is very expensive to be cleared as different rights holders must be identified and then asked to provide all the rights necessary for the work to be usable. The phenomenon of Rights Lowest Common Denominator appears: when multiple parties have rights on the same work, the most restrictive licence terms provided determines the use of the whole work. If no permission is given by just one rights owner, the work cannot be used at all. On the contrary, when the work is used and copies even illegally on the basis of informal copynorms, as is in the case of BlueGreece, the potentials for further creative use and documentation of the works is amplified (Schultz 2007).

### 5.4.1 The problem of Rights Lowest Common Denominator
The conditions of use of an object that comprises multiple layers of rights is set by the lowest common set of rights awarded by all contributors. If a particular owner cannot be identified or refuses permission, the work cannot be legally used (Sterling 2003).

### 5.5 Physical and virtual embodiments of content
It is advisable to differentiate between physical and digital copies of the work as they are governed by different business models (Tsiavos 2006). Also, when a work is digitised, new rights on the digital record are sometimes created. This element of rights creation from physical property has a seemingly paradoxical result: works that are no longer in copyright are more likely to be digitised and exploited as they have lower (or zero) clearance transaction costs. Also, in experience-intensive environments such as museums, the proliferation and free dissemination of digital copies of the work are increasing the value of the original physical object that is more likely to be visited and possibly create revenue for the memory institution. For instance, the digital collection of the BL and BBC archives attracts visitors to the physical space of both institutions.

### 5.5.1 Conclusion
The less rights existing in a work the more likely it is to produce value of any kind as the presence of un- cleared rights radically increases transaction costs.

### 5.6 Maturity of IPR management models
It is neither possible nor desirable to always use a clean hands model. Pure clean hands models are only used in the case where the organisation is only aggregating content that is both licensed and stored by the content providers themselves, such as in the case of the CenturyShare project. In the case of BlueGreece the site only manages links to content and meta-data whereas the actual content is stored by the users. In all other cases, such as in the NLH LOR, the content is centrally stored but directly licensed between the participants of the project. Hybrid models are necessary for securing control points and managing the flows of value in relation to flows of rights and works. The maturity of the IPR management model that allows a project to adopt one or another flow model, depends on the existence of proper IPR documentation, coherent IPR policies and appropriate risk management processes in place. Standardised tools such as the SCA IPR Toolkit could greatly assist organisations or projects that seek to adopt one or another flow model.

### 5.6.1 Conclusion
The type of the IPR management scheme used by an organisation may be assessed on the basis of the existence of IPR documentation, IPR policies and IPR risk management in place and the way they may be serving flows of value. There is need for a Capability Maturity Model for Open Content (Paulk 1995).

5.7 Documentation of layers of rights

The documentation of layers of rights needs to be conducted in a way that is interoperable and transferable (we need to all be using rights management systems that are compatible). In the same way as the sharing of user generated metadata decreases the costs of search for relevant content, the establishment of interoperable rights documentation scheme among SCA sponsor organisations could significantly decrease rights clearance costs.

5.8 The issue of attribution and provenance

The case studies indicate that the more permissions are conferred to the end-user in relation to the distributed content, the more likely it is that attribution and provenance requirements will appear. The reason is that the flows of value that are contingent upon the visibility of the work are non-monetary and mainly have to do with reputation. For example, in the case of Internet Archive, where Creative Commons licences are used allowing users to freely share and in some cases repurpose content, the project provides software for proper attribution or listing of the sources of a derivative work.

When the value also derives from the ability of other users to complement or repurpose the work, it is also necessary to be able to trace contributors both in order to be able to properly attribute and to define collective ownership or even be able to trace potential violations of copyright and/or related rights, such as moral rights or communicate with the author of the repurposed item for further collaboration. This has been experienced in the NLH LOR case.

Even in cases where the objective is not obtaining value, the requirements of attribution and provenance relate to the need to reduce potential costs: in the BBC and NLH LOR project, the main concern with repurposed work is its quality and the need to differentiate user-generated from in-house produced content in order not to harm the institution's reputation. In the BlueGreece case, the objective is again quality, irrespective of the flows of copyrights over the works. This is enforced through community checking and active moderators and is closely monitored by the community.

5.8.1 Conclusion

The closer we get to the model of unrestricted sharing and repurposing of content, the more likely the need for attribution, quality assurance, source tracing and provenance.

5.9 Legal and regulatory issues

The existence of different types of licences for the items stored in different collections requires some sort of licence management system that ranges from simple Excel databases (as used in the BL ASR I project) to the SPECTRUM standard used by the Collections Trust.

The problem of high clearance costs appears mostly in collections of great cultural but low market value or extensive collections consisting of work with multiple layers of rights (eg in the case of the BL ASR project). In particular:

\*      Large public organisations are obvious litigation targets, they are difficult to be indemnified and run great reputation risks from violating any IPR-related rules

\*      The economic rationale behind the existing copyright laws is appropriate for works that have a clear market value, such as commercial sound recordings (Gordon 1989; Boyle 1992; Barlow 1994; Boyle 1996; Ghosh 1998; Benkler 1999; Watt 2000). However, it is inappropriate for works with low market value, and often not properly documented, but with high cultural and educational value. For such works the costs of identification and negotiations of rights is far greater than the actual cost of acquiring the rights. Such costs often cancel any effort to make them available. This is the case with orphan works (Brito and Dooling 2005; Huang 2006; Boyle 2008) and has been very vocally expressed in the case of the BL ASR collections

\*      When a work comprises multiple layers of rights belonging to more than one rights holder, it is most likely that the transaction costs of clearance will make its digitisation or dissemination impractical. This is not merely a result of the primary costs described in the previous points but also due to the incremental cost that each additional work has for the whole of the project in terms of time: any publicly funded project has to be completed within a certain time frame and this is not possible if the rights are not previously cleared. The situation is extremely difficult: the funding is for content that will be made publicly available but the content cannot be made available if they are not cleared. If the content is first cleared and then digitised, then the risk of project delay appears as clearance procedures can be extremely lengthy. If the content is first digitised and then cleared, then the project runs the risk of having digitised material that will never appear in public. This might be in breach of the funding agreement, and certainly will involve wasted time and money. These problems appear in particular in the BL ASR project.

\* The optimal regulatory mixture takes into account a combination of legal (licensing based in particular) and technological means (Black 2000; Black 2001; Murray and Scott 2002; Wu 2003; Lessig 2006; Murray 2007).

5.9.1 Conclusion
The 'IPR jam' or 'licence pollution' phenomenon describes the situation where existence of multiple layers of rights and rights holders on a single object make any extraction of value impossible (Elkin-Koren 1997; Elkin-Koren 1998; Elkin-Koren 2005; Elkin-Koren 2006).